\documentclass[floats, eqnum, showpacs, nofootinbib, 
preprint,
eqsecnum]{revtex4-1}

\usepackage{color,graphicx}
\usepackage{amsfonts}
\usepackage{amssymb}

\begin{document}

\title{Particle collision with an arbitrarily high center-of-mass energy near a Ba\~{n}ados-Teitelboim-Zanelli black hole}
\author{Naoki Tsukamoto${}^{1}$}\email{tsukamoto@rikkyo.ac.jp}
\author{Kota Ogasawara${}^{2}$}\email{k.ogasawara@rikkyo.ac.jp}
\author{Yungui Gong${}^{1}$}\email{yggong@hust.edu.cn}
\affiliation{
${}^{1}$School of Physics, Huazhong University of Science and Technology, Wuhan 430074, China \\
${}^{2}$Department of Physics, Rikkyo University, Tokyo 171-8501, Japan 
}

\begin{abstract}
We consider a particle collision with a high center-of-mass energy near a Ba\~{n}ados-Teitelboim-Zanelli (BTZ) black hole.
We obtain the center-of-mass energy of two general colliding geodesic particles in the BTZ black hole spacetime. 
We show that the center-of-mass energy of two ingoing particles can be arbitrarily large on an event horizon
if either of the two particles has a critical angular momentum and the other has a non-critical angular momentum.
We also show that the motion of a particle with a subcritical angular momentum is allowed near an extremal rotating BTZ black hole 
and that a center-of-mass energy for a tail-on collision at a point can be arbitrarily large in a critical angular momentum limit.
\end{abstract}

\preprint{RUP-17-10}

\maketitle

\section{Introduction}
Recently, LIGO reported 
three gravitational-wave events GW150914~\cite{Abbott:2016blz}, 
GW151226~\cite{Abbott:2016nmj}, and GW170104~\cite{Abbott:2017vtc} and they showed that stellar-mass black holes exist in nature.
Phenomena in a strong gravitational field near black holes become a more important topic not only in general relativity but also in astronomy and astrophysics.
The black holes in nature will be described well as the Kerr black hole solution which is the vacuum solution of the Einstein equations.

In 2009, Ba\~{n}ados, Silk, and West refound that 
the center-of-mass energy of two colliding particles near an event horizon can be arbitrarily large in the extremal Kerr black hole spacetime
if either of the two ingoing particles has a critical angular momentum and the other does not have the critical angular momentum~\cite{Banados:2009pr}. 
The particle collision is often called Ba\~{n}ados-Silk-West (BSW) collision or BSW process 
but we should mention that the particle collision with the infinite center-of-mass energy was pointed out by Piran, Shaham, and Katz in 1975~\cite{Piran_1975}. 
In Ref.~\cite{Banados:2009pr} an on-equatorial-plane collision was considered 
but collisions with an arbitrarily high center-of-mass energy occur also off the equatorial plane~\cite{Harada:2011xz}.
The effect of a weak electromagnetic field on the BSW collision~\cite{Igata:2012js} and the BSW collision in the near-horizon geometry of the extremal Kerr black hole spacetime were discussed~\cite{Galajinsky:2013as}.
Two ingoing neutral particle collide with an arbitrarily large center-of-mass energy in a near horizon limit not only in the extremal Kerr black hole spacetime but also
in the Kerr-Newmann spacetime~\cite{Wei:2010vca}, in the Kerr naked singularity spacetime~\cite{Patil:2011ya} 
in the Kerr-de Sitter black hole spacetime~\cite{Li:2010ej},
in higher dimensional black hole spacetimes~\cite{Abdujabbarov:2013qka,Tsukamoto:2013dna},
and in lower dimensional black hole spacetimes~\cite{Lake:2010bq,Yang:2012we,Sadeghi:2013gmf,Fernando:2017kut}.

After rediscovering of the BSW process, several aspects of the BSW process were criticized by several authors~\cite{Berti:2009bk,Jacobson:2009zg}.
For the infinite center-of-mass energy, the angular momentum of either of two particles must be fine-tuned to be a critical value. 
The particle with the critical angular momentum rotates around the extremal Kerr black hole infinite times 
and then reach an event horizon in infinite proper time.
Even if extremal Kerr black holes exist in nature, the backreaction of gravitational waves will significantly affect the BSW process.

The high center-of-mass energy of two colliding particles near an event horizon does not mean 
that observers at infinity obtain high energy particles or massive particles. 
Creations after the BSW collision near the event horizon are strongly red-shifted and the escape fraction can be diminished~\cite{McWilliams:2012nx}. 
On the other hand, the BSW collision stimulates to reconsider the details of a collisional Penrose process 
which is a process of extraction of energy from a rotating Kerr black hole after a particle collision~\cite{Piran:1977dm,Bejger:2012yb,Harada:2012ap,Schnittman:2014zsa,Berti:2014lva,Leiderschneider:2015ika,Leiderschneider:2015kwa,Ogasawara:2015umo,Harada:2016eff,Ogasawara:2016yfk}.

In Ref.~\cite{Zaslavskii:2010aw}, 
Zaslavskii found the electromagnetic counterpart of the BSW collision in the extremal Reissner-Nordstr\"{o}m black hole spacetime.
A particle collision with an arbitrarily high center-of-mass energy occurs 
when either of two ingoing particles has a critical charge and when the other does not have the critical charge.
The critical charged particle reaches in arbitrarily long proper time to an extremal event horizon.
The electromagnetic counterpart of the BSW collision was also found in a higher-dimensional extremal charged black hole spacetime~\cite{Tsukamoto:2013dna}.
The BSW effect will be a universal phenomena in near-extremal and extremal spacetimes with and without an event horizon.
Simple kinematic explanations of the BSW collision were given in Refs.~\cite{Jacobson:2009zg,Zaslavskii:2011dz}.
In Ref.~\cite{Tsukamoto:2013dna}, a tight link between the BSW collision and a test-field instability of an extremal horizon 
in asymptotically flat and extremal black hole spacetimes~\cite{Aretakis:2011gz,Aretakis:2011hc,Aretakis:2011ha,Aretakis:2012ei,Aretakis:2012bm,Aretakis:2013dpa,Murata:2012ct,Murata:2013daa} 
was pointed out.
See a review~\cite{Harada:2014vka} for the more details of the BSW process. 

The treatment of gravity generated by colliding particles is a problem of the BSW collision~\cite{Kimura:2010qy}. 
Since particles near an extremal event horizon have a large energy, 
the gravity of the particles will affect on the geodesic motion.
Thus, the effect of self-gravity of particles on the BSW collision cannot be neglected.
The analytical treatment of the self-gravity of the particles in the Kerr spacetime is a challenging problem because of low symmetry of the spacetime.
If we treat analytically fast rotating thin shells in the Kerr spacetime,
we can estimate the effect of the self-gravity on the BSW collision.
The analytical description of fast rotating thin shells in the Kerr spacetime, 
however, is also a difficult problem to solve~\cite{Mann:2008rx,Delsate:2014iia,Rocha:2015tda}. 

Considering collisions of two charged thin shells including their self-gravity 
in the Reissner-Nordstr\"{o}m spacetime 
helps us to estimate the effect of gravity generated by colliding particles on the BSW collision~\cite{Kimura:2010qy,Nakao:2013uj}.
Kimura \textit{et al.} showed that the center-of-mass energy of the BSW collision of the two charged shells cannot be arbitrarily large~\cite{Kimura:2010qy}.

Investigating the BSW collision in 2+1 dimension can be another good approach to estimate the effect of self-gravity of particles on the BSW collision
since the collision of fast rotating dust thin shells in 2+1 dimension 
is more tractable than in the Kerr spacetime~\cite{Mann:2008rx,Rocha:2011wp}.

In this paper, we consider a particle collision in the  Ba\~{n}ados-Teitelboim-Zanelli (BTZ) black hole spacetime with an angular momentum 
and a negative cosmological constant~\cite{Banados:1992wn,Banados:1992gq} motivated 
by further investigations for the backreaction effect of the self-gravity of particles on the BSW collision~\cite{Follow-up}.
The BTZ black hole is considered as a typical black hole in 2+1 dimension because of the existence of a no-go theorem 
for asymptotically flat and stationary black holes satisfying the dominant energy condition in 2+1 dimensions in Einstein gravity~\cite{Ida:2000jh}.

One may suspect that the negative cosmological constant affects on the particle collision and it is very different from the BSW collision in the Kerr spacetime.
The effect of the negative cosmological constant on the BSW collision will be negligible since the collision with a high center-of-mass energy 
occurs near an extremal event horizon. 

We concentrate on the collision of two particles on an event horizon and outside of the BTZ black hole in this paper
while a particle collision was considered on the event horizon and inside the BTZ black hole~\cite{Lake:2010bq,Yang:2012we} 
motivated by the internal instability of black holes.

This paper is organized as follows. 
In Sec.~II, we consider a particle motion and the center-of-mass energy of the collision of two particles in the BTZ black hole spacetime.
In Sec.~III, we investigate the motion of particles with critical and subcritical angular momenta near the BTZ black hole.
In Sec.~IV, we investigate a tail-on collision of two particles with subcritical angular momentums near an extremal BTZ black hole.
In Sec.~V, we summarise our result.
In this paper we use the units in which the light speed is unity.

\section{Center-of-mass energy for particle collision in the BTZ black hole spacetime}
In this section, we review a particle motion and investigate the center-of-mass energy of the collision of two particles in the BTZ black hole spacetime.

\subsection{Line element}
The line element in the BTZ black hole spacetime is given by
\begin{equation}\label{eq:line_element1}
ds^{2}=-f(r)dt^{2}+\frac{dr^{2}}{f(r)}+r^{2} \left( d\phi-\frac{4GJ}{r^{2}}dt \right)^{2},
\end{equation}
where
\begin{equation}
f(r)=-8GM+\frac{r^{2}}{l^{2}}+\frac{16G^{2}J^{2}}{r^{2}},
\end{equation}
$G$ is a gravitational constant,
$M$ and $J$ are the Arnowitt-Deser-Misner (ADM) mass and the angular momentum of a BTZ black hole, respectively,
and $l$ is the radius of the curvature related to a negative cosmological constant $\Lambda<0$ by 
\begin{equation}
l\equiv \sqrt{\frac{1}{-\Lambda}}.
\end{equation}
If a condition 
\begin{equation}
M \geq \frac{\left|J\right|}{l}
\end{equation}
is satisfied,
outer and inner horizons exist at $r=r_{+}$ and $r=r_{-}$, respectively, where $r_{\pm}$ is defined as
\begin{equation}
r_{\pm}=2l\sqrt{GM \left( 1\pm \sqrt{1-\frac{J^{2}}{M^{2}l^{2}}} \right) }.
\end{equation}
Please notice that $r_{+} \geq r_{-}$ is satisfied and that the outer horizon is an event horizon. 
BTZ black holes with the maximal angular momentum $\left|J\right|=Ml$ are said to be extremal.
When the BTZ black hole is extremal, the outer and inner horizons are coincide,
\begin{equation}
r_{+}=r_{-}=2l\sqrt{GM}.
\end{equation}
There are time translational and axial Killing vectors, $t^{\mu}\partial_{\mu}=\partial_{t}$ and $\phi^{\mu}\partial_{\mu}=\partial_{\phi}$, 
because of stationarity and axial symmetry of the spacetime, respectively.

Using the radii of the outer and inner horizons $r_{\pm}$, the line element, the function $f(r)$, the ADM mass $M$, and the angular momentum $J$ are expressed as
\begin{equation}\label{eq:line_element2}
ds^{2}=-f(r)dt^{2}+\frac{dr^{2}}{f(r)}+r^{2} \left( d\phi-\frac{r_{+}r_{-}}{lr^{2}}dt \right)^{2},
\end{equation}
\begin{equation}
f(r)=\frac{(r^{2}-r^{2}_{+})(r^{2}-r_{-}^{2})}{l^{2}r^{2}},
\end{equation}
\begin{equation}
M=\frac{r_{+}^{2}+r_{-}^{2}}{8Gl^{2}},
\end{equation}
and
\begin{equation}
J=\frac{r_{+}r_{-}}{4Gl},
\end{equation}
respectively.

\subsection{Particle motion}
We consider the motion of a particle with a three momentum $p^{\mu}$ and a rest mass $m$ in the BTZ black hole spacetime.
The conserved energy of the particle
\begin{equation}\label{eq:E}
E\equiv -g_{\mu\nu}t^{\mu}p^{\nu}=-g_{t\nu}p^{\nu}
\end{equation} 
and the conserved angular momentum of the particle
\begin{equation}\label{eq:L}
L\equiv g_{\mu\nu}\phi^{\mu}p^{\nu}=g_{\phi \nu}p^{\nu}
\end{equation} 
are constant along a geodesic. 
We assume that $E$ is positive.
From the condition $p^{\mu}p_{\mu}=-m^{2}$ and Eqs.~(\ref{eq:E}) and (\ref{eq:L}), 
the components of the three momentum $p^{\mu}$ are obtained as
\begin{eqnarray}
&&p^{t}(r)=\frac{S(r)}{f(r)}, \\\label{eq:pr}
&&p^{r}(r)=\sigma \sqrt{R(r)},\\
&&p^{\phi}(r)=\frac{r_{+}r_{-}S(r)}{lf(r)r^{2}}+\frac{L}{r^{2}},
\end{eqnarray}
where $R(r)$ and $S(r)$ are defined as 
\begin{equation}\label{eq:R}
R(r)\equiv S^{2}(r)- \left( m^{2}+\frac{L^{2}}{r^{2}} \right) f(r)
\end{equation}
and 
\begin{equation}
S(r)\equiv E-\frac{r_{+}r_{-}L}{lr^{2}},
\end{equation}
respectively, and where $\sigma=\pm 1$ is chosen as $+1$ for an outgoing particle and $-1$ for an ingoing particle.
We assume a forward-in-time condition $p^{t}(r)\geq 0$, i.e., $S(r)\geq 0$, for $r \geq r_{+}$. 
We define critical angular momentum $L_{c}$ of a particle as $L_{c}\equiv r_{+}lE/r_{-}$. 
A particle with the critical angular momentum $L=L_{c}$ satisfies a condition $S(r_{+})=0$.
A particle rotates in the $\phi$ direction ($-\phi$ direction) when $p^{\phi}\geq 0$ ($p^{\phi}< 0$) is satisfied.
 
Using $p^{\mu}=dx^{\mu}/d\lambda$, where $\lambda$ is the parameter of the geodesic,
and using Eq.~(\ref{eq:pr}), we obtain the equation of the motion in the radial direction of a particle as
\begin{equation}
\frac{1}{2} \left( \frac{dr}{d\lambda} \right)^{2}+ V(r)=0, 
\end{equation}
where $V(r)$ is an effective potential in the radial direction of the particle defined by $V(r)\equiv -R(r)/2$.
The motion of the particle is allowed in regions where $V(r)\leq 0$ or $R(r)\geq 0$
while it is prohibited in regions where $V(r)>0$ or $R(r)<0$.
The particle can exist on the event horizon $r=r_{+}$ because of $R(r_{+})=S^{2}(r_{+})\geq 0$. 

If a particle has a mass, it cannot exist at infinity because of
\begin{equation}\label{eq:R_infty}
\lim_{r \rightarrow \infty}R(r)=\lim_{r\rightarrow \infty} -\frac{m^{2}r^{2}}{l^{2}}<0.
\end{equation}

For a massless particle, we obtain $R(r)$ as 
\begin{equation}
R(r)= E^{2}-\frac{L^{2}}{l^{2}} + \frac{L}{lr^{2}}\left[ -2Er_{+}r_{-}+\frac{L}{l}(r_{+}^{2}+r_{-}^{2}) \right]. 
\end{equation}
If $-El\leq L \leq El$ is satisfied, $R(r)$ is nonnegative in the range of $r_{+} \leq r$. 
If $L < -El$ or $El < L \leq L_{c}$ is satisfied, $R(r)$ is nonnegative in the range of $r_{+} \leq r \leq r_{0}$ 
and $R(r)$ is negative in the range of $r_{0}<r$, 
where $r_{0}$ is given by 
\begin{equation}
r_{0}\equiv r_{+}  \sqrt{ 1- \frac{l^{2}S^{2}(r_{+})}{E^{2}l^{2}-L^{2}} }.
\end{equation}
Notice $r_{+}<r_{0}$ for $L < -El$ or $El < L \leq L_{c}$.

\subsection{Static limit, ergo region, and Penrose process}
The $(t,t)$ component of the metric tensor in the BTZ black hole spacetime becomes nonpositive, 
i.e., $g_{tt}(r) \leq 0$, for a region between the event horizon $r=r_{+}$ and a radius $r=r_{\mathrm{sl}}$,
where $r_{\mathrm{sl}}$ is defined as
\begin{equation}
r_{\mathrm{sl}}\equiv \sqrt{r_{+}^{2}+r_{-}^{2}}.
\end{equation}

Since the time translational Killing vector $t^{\mu}$ is spacelike in the region $r_{+} \leq  r \leq r_{\mathrm{sl}}$,
static observers with a three velocity proportional to the time translational Killing vector cannot exist in there. 
We call the radius $r=r_{\mathrm{sl}}$ static limit and call the region ergo region. 
Inside of the ergo region, the conserved energy $E$ of a particle can be nonpositive.

Consider a particle with a positive conserved energy comes from infinity and enters into an ergo region and decays into two particles in there.
A production with a negative conserved energy falls into the rotating black hole and another production 
with a larger conserved energy than the conserved energy of the incident particle escapes infinity.
This is a Penrose process known as the process of the extraction of an energy from a rotating black hole~\cite{Penrose:1969pc}.
Cruz \textit{et al.} considered the Penrose process in BTZ black hole spacetime~\cite{Cruz:1994ir}.
They concluded that the Penrose process does not occur if the emitted particle has a mass because it cannot reach into infinity
and that the Penrose process does occur if a massless particle is emitted.

\subsection{Center-of-mass energy for a particle collision} 
We consider the collision of two particles which are named particle $1$ and $2$ in the BTZ black hole spacetime.
We obtain the formula of the center-of-mass energy $E_{CM}(r)$ of the two general geodesic particles at a collisional point in the BTZ black hole spacetime as 
\begin{eqnarray}
E_{CM}^{2}(r)
&\equiv& -(p_{1}^{\mu}(r)+p_{2}^{\mu}(r))(p_{1 \mu}(r)+p_{2 \mu}(r))\nonumber\\
&=&m_{1}^{2}+m_{2}^{2}
+2 \frac{S_{1}(r)S_{2}(r)-\sigma_{1}\sigma_{2}\sqrt{R_{1}(r)R_{2}(r)}}{f(r)} \nonumber\\
&&-2\frac{L_{1}L_{2}}{r^{2}},  
\end{eqnarray}
where $R_{k}(r)$ and $S_{k}(r)$ are defined by
\begin{equation}
R_{k}(r)\equiv S_{k}^{2}(r)- \left( m_{k}^{2}+\frac{L_{k}^{2}}{r^{2}} \right) f(r)
\end{equation}
and 
\begin{equation}
S_{k}(r)\equiv E_{k}-\frac{r_{+}r_{-}L_{k}}{lr^{2}},
\end{equation}
respectively, and where $p_{k}^{\mu}$, $m_{k}$, $E_{k}$, $L_{k}$, and $\sigma_{k}$ 
denote $p^{\mu}$, $m$, $E$, $L$, and $\sigma$ of particle $k$, respectively, where $k=1$ and $2$.~\footnote{
In Refs.~\cite{Lake:2010bq,Yang:2012we,Hussain:2012su}, an equal mass $m_{1}=m_{2}$ was assumed while it is not assumed in this paper.
In Ref.~\cite{Hussain:2012su}, $L_{1}=L_{2}=0$ was also assumed.  
}

We consider the collision of two ingoing particles. 
In this case, we should set $\sigma_{1}=\sigma_{2}=-1$. 
The center-of-mass energy of the tail-on collision is given by
\begin{eqnarray}
E_{CM}^{2}(r)
&=&m_{1}^{2}+m_{2}^{2}
+2 \frac{S_{1}(r)S_{2}(r)-\sqrt{R_{1}(r)R_{2}(r)}}{f(r)} \nonumber\\
&&-2\frac{L_{1}L_{2}}{r^{2}}.
\end{eqnarray}

Both the numerator and the denominator of the third term vanish in the near horizon limit $r\rightarrow r_{+}$.
Using l'Hopital's rule with respect to $r$, the third term becomes in the near horizon limit $r\rightarrow r_{+}$ 
\begin{eqnarray}
&&\lim_{r \rightarrow r_{+}} 2 \frac{S_{1}(r)S_{2}(r)-\sqrt{R_{1}(r)R_{2}(r)}}{f(r)} \nonumber\\
&=&\frac{S_{2}(r_{+})}{S_{1}(r_{+})} \left( \frac{L_{1}^{2}}{r^{2}_{+}}+m_{1}^{2} \right)
+\frac{S_{1}(r_{+})}{S_{2}(r_{+})} \left( \frac{L_{2}^{2}}{r^{2}_{+}}+m_{2}^{2} \right). \nonumber\\
\end{eqnarray}
Thus, the center-of-mass energy of the two ingoing particles in the near horizon limit $r\rightarrow r_{+}$ is obtained as
\begin{eqnarray}
\lim_{r \rightarrow r_{+}} E_{CM}^{2}(r)
&=&m_{1}^{2}+m_{2}^{2}
+\frac{S_{2}(r_{+})}{S_{1}(r_{+})} \left( \frac{L_{1}^{2}}{r^{2}_{+}}+m_{1}^{2} \right) \nonumber\\
&&+\frac{S_{1}(r_{+})}{S_{2}(r_{+})} \left( \frac{L_{2}^{2}}{r^{2}_{+}}+m_{2}^{2} \right) 
-2\frac{L_{1}L_{2}}{r_{+}^{2}}. \nonumber\\
\end{eqnarray}
This shows that the center-of-mass energy of the tail-on collision in the near horizon limit $r\rightarrow r_{+}$ in the BTZ black hole spacetime 
can be arbitrarily large if and only if the one of the particles has the critical angular momentum $L=L_{c}$ 
and the other has a non-critical angular momentum.

If both the two particles have the critical angular momentum, i.e., $L_{1}=L_{c1}$ and $L_{2}=L_{c2}$, where $L_{c1}$ and $L_{c2}$ are 
the critical angular momenta for particle $1$ and $2$, respectively, 
the center-of-mass energy in the near horizon limit $r\rightarrow r_{+}$ becomes
\begin{equation}
\lim_{r \rightarrow r_{+}} E_{CM}^{2}(r)
= \left( 1 +\frac{E_{2}}{E_{1}} \right) m_{1}^{2} +\left( 1 +\frac{E_{1}}{E_{2}} \right) m_{2}^{2}.
\end{equation}

\section{Motion of a particle with the critical and subcritical angular momentum}
In this section, we show that a particle with the critical angular momentum cannot exist on the outside of the BTZ black hole 
while a particle with a subcritical angular momentum can exist on the event horizon and outside of the BTZ black hole.

\subsection{Motion of a particle with the critical angular momentum}
We investigate the motion of a particle with the critical angular momentum.
The components of the three momentum of the particle with critical angular momentum are given by
\begin{eqnarray}
&&p^{t}(r)=\frac{El^{2}}{r^{2}-r_{-}^{2}}, \\
&&p^{r}(r)=\sigma \sqrt{R(r)}, \\
&&p^{\phi}(r)=\frac{Elr_{+}}{(r^{2}-r_{-}^{2})r_{-}},
\end{eqnarray}
where 
\begin{eqnarray}\label{eq:R_c}
R(r)
&=&R_{c}(r)\nonumber\\
&\equiv&-\frac{r^{2}-r_{+}^{2}}{r^{2}}\left[ \frac{E^{2}(r_{+}^{2}-r_{-}^{2})}{r_{-}^{2}}+\frac{m^{2}(r^{2}-r_{-}^{2})}{l^{2}} \right]. \nonumber\\
\end{eqnarray}
Equation~(\ref{eq:R_c}) shows that $R_{c}(r)$ vanishes on the event horizon $r=r_{+}$ 
and that $R_{c}(r)$ is negative outside of the event horizon $r>r_{+}$.
Thus, the particle with the critical angular momentum cannot exist outside of the event horizon $r>r_{+}$.

The derivative of $R_{c}(r)$ with respective to $r$ is given by 
\begin{equation}\label{eq:R_c'}
R_{c}'(r)=-\frac{2m^{2}r}{l^{2}}+\frac{2r_{+}^{2}}{r^{3}r_{-}^{2}l^{2}}\left[ -E^{2}l^{2}(r_{+}^{2}-r_{-}^{2})+r_{-}^{4}m^{2} \right],
\end{equation}
where $'$ denotes the differentiation with respect to $r$
and it becomes, on the event horizon $r=r_{+}$,
\begin{eqnarray}\label{eq:R_c'+}
R_{c}'(r_{+})=-\frac{2(r_{+}^{2}-r_{-}^{2})(l^{2}E^{2}+r_{-}^{2}m^{2})}{r_{+}r_{-}^{2}l^{2}}\leq 0.
\end{eqnarray}

In the extremal case, i.e., $r_{+}=r_{-}$, we obtain  
\begin{equation}\label{eq:R_c_extremal}
R_{c}(r)=-\frac{m^{2}(r^{2}-r_{+}^{2})^{2}}{r^{2}l^{2}}
\end{equation}
and 
\begin{equation}\label{eq:R'_c_extremal}
R'_{c}(r)=-\frac{2m^{2}(r^{2}-r_{+}^{2})(r^{2}+r_{+}^{2})}{r^{3}l^{2}}.
\end{equation}
Thus, we get $R_{c}(r_{+})=R_{c}'(r_{+})=0$ on the event horizon.

\subsection{Motion of a particle with a subcritical angular momentum}
We consider the motion of a particle with a subcritical angular momentum $L=L_{c} - r_{+}l \delta /r_{-}=r_{+}l (E-\delta)/r_{-}$, 
where $\delta$ is a positive constant.
From Eq.~(\ref{eq:R}), $R(r)$ is obtained as 
\begin{equation}
R(r)=R_{c}(r)+\frac{r_{+}^{2}\delta \left[ 2(r^{2}-r_{+}^{2})E+(-r^{2}+r_{+}^{2}+r_{-}^{2})\delta \right]}{r_{-}^{2}r^{2}}.
\end{equation}
The particle with the subcritical angular momentum can exist on the event horizon and nearly outside of the black hole
since
\begin{equation}\label{eq:R_sub_horizon}
R(r_{+})=\delta^{2} > 0.
\end{equation}
The derivative of $R(r)$ with respective to $r$ is obtained as
\begin{equation}\label{eq:R'sub}
R'(r)=R'_{c}(r)+\frac{2r_{+}^{2}\delta \left[ 2r_{+}^{2}E-(r_{+}^{2}+r_{-}^{2})\delta \right]}{r_{-}^{2}r^{3}}
\end{equation}
and it becomes, on the horizon, 
\begin{equation}\label{eq:R'sub_horizon}
R'(r_{+})=R'_{c}(r_{+})+\frac{ 4r_{+}^{2}E\delta -2(r_{+}^{2}+r_{-}^{2})\delta^{2}}{r_{-}^{2}r_{+}}.
\end{equation}

\begin{figure}[htbp]
\begin{center}
\includegraphics[width=87mm]{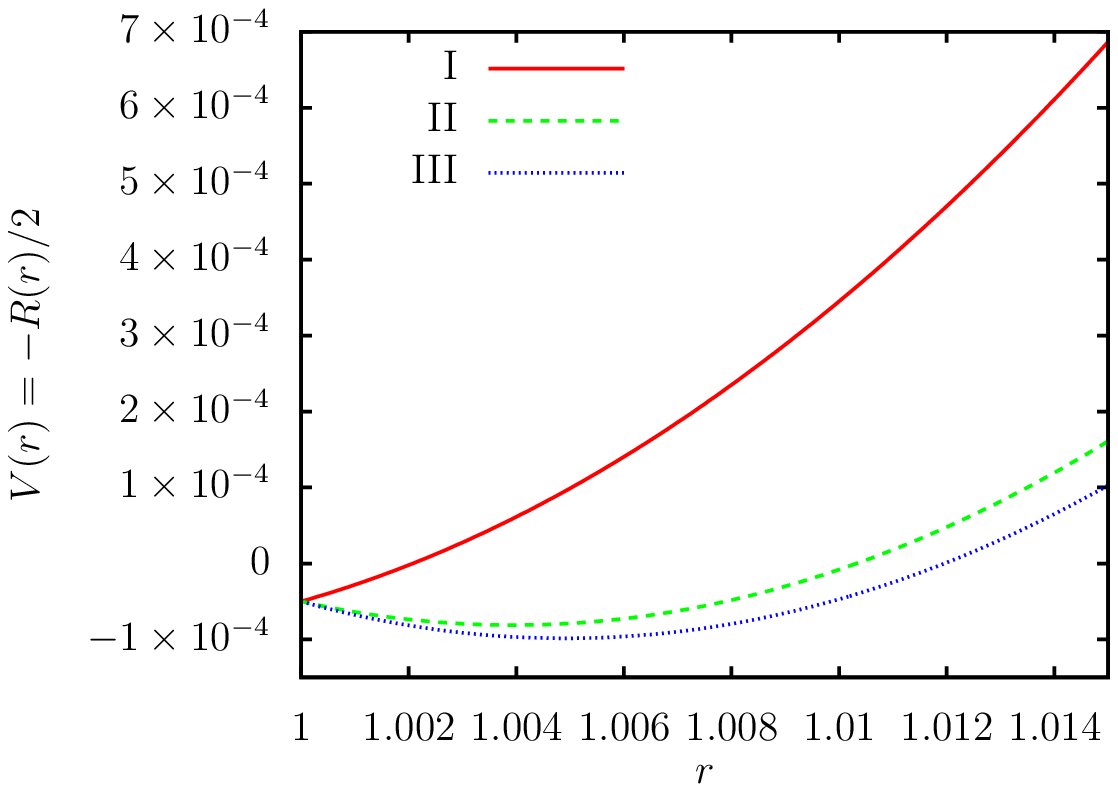}
\end{center}
\caption{The examples of the effective potential $V(r)=-R(r)/2$ of the radial motion of a particle 
with $m=E=1$ and $\delta=10^{-2}$ in the BTZ black hole spacetime with $r_{+}=l=1$. 
The solid~(red), dashed~(green), and dotted~(blue) curves denote
the effective potential in the cases~I ($r_{-}=0.99$; $V'(r_{+})>0$), II ($r_{-}=0.999$;$V'(r_{+})<0$), and III ($r_{-}=1$; $V'(r_{+})<0$), respectively.
Notice that the spacetime is not extremal in the cases I and II and it is extremal in the case III.}
\label{fig:Vb}
\end{figure}

From Eqs.~(\ref{eq:R_c'+}) and (\ref{eq:R'sub_horizon}), 
$R'(r_{+})$ is a quadratic equation with respect to $\delta$. 
After some straightforward calculation, we obtain the following results.
When the conserved energy $E$ is smaller than $E_{m}$ defined as
\begin{equation}
E_{m} \equiv  \frac{m \sqrt{r_{+}^{4}-r_{-}^{4}}}{lr_{-}}, 
\end{equation}
$R'(r_{+})$ is negative.
When the conserved energy $E$ is larger than or is equal with $E_{m}$,
$R'(r_{+})$ is nonnegative if and only if $\delta_{L} \leq \delta \leq \delta_{R}$ is satisfied.
Here $\delta_{L}$ and $\delta_{R}$ are defined as
\begin{equation}
\delta_{L} \equiv \frac{r_{+}^{2}El-r_{-}\sqrt{r_{-}^{2}E^{2}l^{2}-(r_{+}^{4}-r_{-}^{4})m^{2}}}{l(r_{+}^{2}+r_{-}^{2})}
\end{equation}
and
\begin{equation}
\delta_{R} \equiv \frac{r_{+}^{2}El+r_{-}\sqrt{r_{-}^{2}E^{2}l^{2}-(r_{+}^{4}-r_{-}^{4})m^{2}}}{l(r_{+}^{2}+r_{-}^{2})},
\end{equation}
respectively.
We notice that an inequality $0 \leq \delta_{L} \leq \delta_{R} \leq E$ is satisfied from the definitions.
Figure~\ref{fig:Vb} shows the effective potential $V(r)=-R(r)/2$ of the radial motion of the particle 
with the subcritical angular momentum in the BTZ black hole spacetime. 

If $E \geq E_{m}$ and $\delta_{L} \leq \delta \leq \delta_{R}$ are satisfied, $R'(r)=0$ has only two real solutions. 
Either of the two solutions is positive and the other solution is negative. 
The positive solution is given by $r=r_{m}$, where 
\begin{equation}
r_{m} \equiv  \left( 1+\frac{R'(r_{+})l^{2}}{2r_{+}m^{2}} \right)^{\frac{1}{4}} r_{+} \geq r_{+}.
\end{equation}
Since $R(r)$ is continuous in the range of $r\geq r_{+}$ 
and since $R(r_{+})=\delta^{2} > 0$, $R'(r_{+}) \geq 0$, $\lim_{r \rightarrow \infty}R(r)<0$, and $\lim_{r \rightarrow \infty}R'(r)<0$, 
$R(r_{m})$ gives the maximal value of $R(r)$ in the range of $r\geq r_{+}$.
The maximal value $R(r_{m})$ is larger than $\delta^{2}$.
Please notice that the effective potential $V(r)=-R(r)/2$ of the radial motion of the particle 
with the subcritical angular momentum takes a minimal value at $r=r_{m}$ in the range of $r\geq r_{+}$.  

\begin{figure}[htbp]
\begin{center}
\includegraphics[width=87mm]{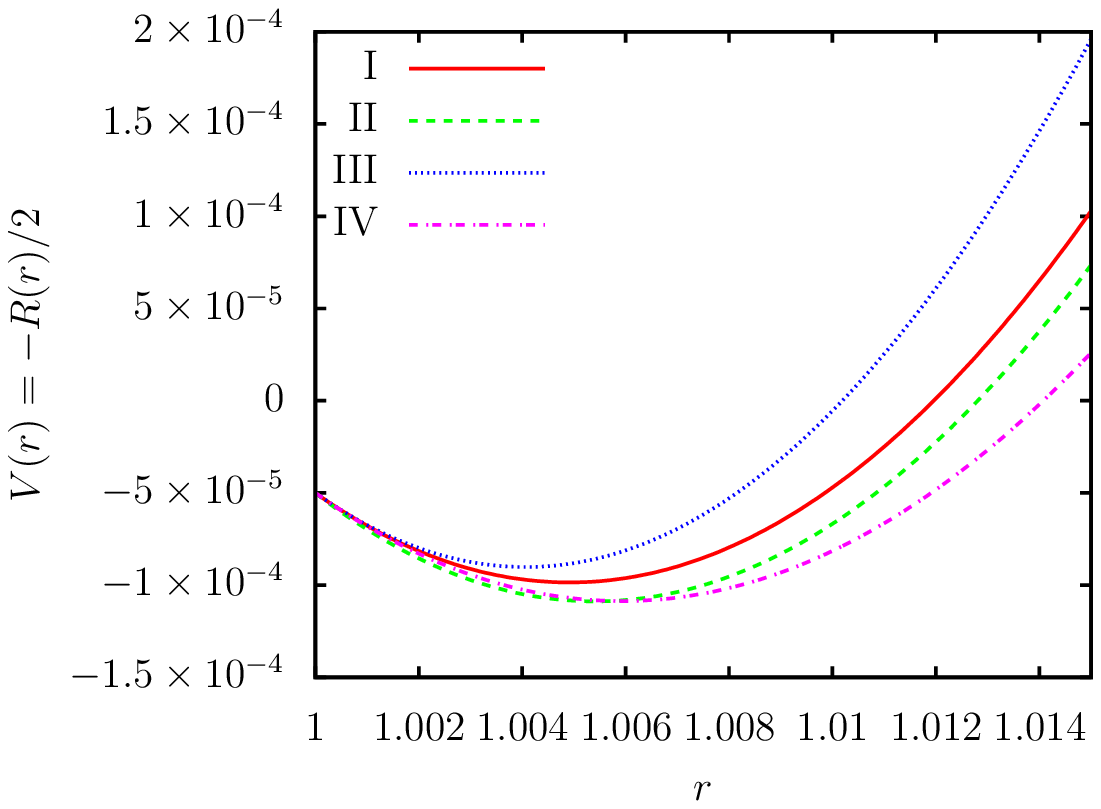}
\end{center}
\caption{The examples of the effective potential $V(r)=-R(r)/2$ in the extremal case. 
The solid~(red), dashed~(green), dotted~(blue) and dash-dotted~(purple) curves denote
the effective potential for the particle with a subcritical angular momentum with $\delta=10^{-2}$ 
in the cases~I ($r_{+}=r_{-}=E=m=l=1$), II ($r_{+}=r_{-}=m=l=1$ and $E=1.1$), III ($r_{+}=r_{-}=E=l=1$ and $m=1.1$), 
and IV ($r_{+}=r_{-}=E=m=1$ and $l=1.1$), respectively.
}
\label{fig:V}
\end{figure}

Figure~\ref{fig:V} shows the effective potential $V(r)=-R(r)/2$ in the extremal spacetime, i.e., $r_{+}=r_{-}$. 
In the extremal case, we obtain $E_{m}=0$, $\delta_{L}=0$, and $\delta_{R}=E$.

In the critical angular momentum limit $\delta \rightarrow 0$, 
we obtain $r_{m} \rightarrow r_{+}$, $R(r_{m})\rightarrow  0$, and $R'(r_{m})\rightarrow  0$ 
in the extremal BTZ black hole spacetime.

\section{A particle collision near an extremal BTZ black hole and its critical angular momentum limit}
Let us assume that a BTZ black hole is extremal,
that particle $1$ has a subcritical angular momentum $L_{1}=l (E_{1}-\delta_{1})\leq L_{c1}$,  
that $\delta_{1}$ satisfies $0=E_{m1}=\delta_{L1} \leq \delta_{1} \leq \delta_{R1}=E_{1}$,
and that particle $2$ has a subcritical angular momentum $L_{2}\leq L_{c2}$.
Here $\delta_{1}$, $E_{m1}$, $\delta_{L1}$, and $\delta_{R1}$ are $\delta$, $E_{m}$, $\delta_{L}$, and $\delta_{R}$ for particle $1$, respectively.
We consider the collision of the two ingoing particles at $r=r_{m}$ 
where is the point of the minimum of the effective potential for the radial motion of particle $1$ in the range of $r\geq r_{+}$.   
The center-of-mass energy of the tail-on collision at $r=r_{m}$ in the critical angular momentum limit $\delta_{1} \rightarrow 0$ is given by
\begin{equation}
\lim_{\delta_{1}\rightarrow 0} E_{CM}^{2}(r_{m})
= m_{1}^{2}+m_{2}^{2} +T -\frac{2lE_{1}L_{2}}{r_{+}^{2}},
\end{equation}
where $T$ is defined by
\begin{equation}
T\equiv \lim_{\delta_{1}\rightarrow 0} 2 \frac{S_{1}(r_{m})S_{2}(r_{m})-\sqrt{R_{1}(r_{m})R_{2}(r_{m})}}{f(r_{m})}.
\end{equation}
The numerator and the denominator of $T$ vanish in the critical angular momentum limit $\delta_{1} \rightarrow 0$.
Using l'Hopital's rule with respect to $\delta_{1}$, $T$ is expressed as
\begin{equation}
T=\lim_{\delta_{1}\rightarrow 0}  \frac{S_{2}(r_{m})}{\dot{f}(r_{m})} \left( 2\dot{S}_{1}(r_{m})-\frac{\dot{R}_{1}(r_{m})}{\sqrt{R_{1}(r_{m})}} \right),
\end{equation}
where $\dot{\,}$ denotes the differentiation with respect to $\delta_{1}$.
Since the numerator and the denominator of $\lim_{\delta_{1}\rightarrow 0} \dot{R}_{1}(r_{m})/\sqrt{R_{1}(r_{m})}$ vanish,
we use l'Hopital's rule with respect to $\delta_{1}$ and we obtain, after a straight-forward calculation,
\begin{equation}
\lim_{\delta_{1}\rightarrow 0}  \frac{\dot{R}_{1}(r_{m})}{\sqrt{R_{1}(r_{m})}} 
=\lim_{\delta_{1}\rightarrow 0}  \sqrt{2\ddot{R}_{1}(r_{m})}
=\lim_{\delta_{1}\rightarrow 0}  2\sqrt{\dot{S}_{1}(r_{m})},
\end{equation}
where 
\begin{equation}
\lim_{\delta_{1}\rightarrow 0} \dot{S}_{1}(r_{m}) = 1 +\frac{l^{2}E_{1}^{2}}{m_{1}^{2}r_{+}^{2}}.
\end{equation}
Thus, $T$ is obtained as
\begin{equation}
T=\lim_{\delta_{1}\rightarrow 0} \frac{S_{2}(r_{m}) \left( \dot{S}_{1}(r_{m})-\sqrt{\dot{S}_{1}(r_{m})} \right)m_{1}^{4}r_{+}^{2}}{l^{2}E_{1}^{2}\delta_{1}}.
\end{equation}
Therefore, the center-of-mass energy $E_{CM}(r_{m})$ 
of the tail-on collision diverges at the point $r=r_{m}$ in the critical angular momentum limit $\delta_{1} \rightarrow 0$.

\section{Summary}
In this paper, we have investigated a collision of two particles in the BTZ black hole spacetime with an angular momentum 
and a negative cosmological constant in 2+1 dimension.
We have obtained a general formula for the center-of-mass energy of two geodesic particles in the BTZ black hole spacetime. 
We have showed that the center-of-mass energy of two ingoing particles in the near horizon limit can be arbitrarily large 
if either of the two particles has a critical angular momentum and the other has a non-critical angular momentum.
We have showed that the motion of a particle with a subcritical angular momentum is allowed near an extremal rotating BTZ black hole 
and that the center-of-mass energy for a tail-on collision at a point near the extremal BTZ black hole can be arbitrarily large 
in the critical angular momentum limit.

Since colliding particles near an event horizon have large energy, 
the self-gravity of the particles affects on the collision.
The center-of-mass energy of the particle collision is suppressed to be a finite value
but the analytical treatment is very difficult.
Since the collision of fast rotating dust thin shells including the self-gravity in the BTZ black hole is more tractable than in the Kerr spacetime~\cite{Mann:2008rx},
we can estimate the backreaction effect of the gravity produced by particles on the BSW collision in the BTZ black hole spacetime from the collision of thin shells.
The collision of two rotating dust thin shells will be developed in a follow-up publication~\cite{Follow-up}.

\section*{Acknowledgements}
The authors would like to show our greatest appreciation to Norihiro Tanahashi for valuable comments and discussions. 
They also thank Masashi Kimura, Jorge V. Rocha, Takafumi Kokubu, Ibrar Hussain, Alexei Deriglazov, Shaoqi Hou, Rio Saitou, and Kumar Shwetketu Virbhadra  
for valuable comments and discussions.
This research was supported in part by the National Natural Science Foundation of China under Grant No. 11475065,
the Major Program of the National Natural Science Foundation of China under Grant No. 11690021.
N.~T. and K.~O. thank the Yukawa Institute for Theoretical Physics at Kyoto University, 
where this work was initiated during the YITP-X-16-10 on "Workshop on gravity and cosmology for young researchers" 
supported by the MEXT KAKENHI No. 15H05888.

\end{document}